# Empirical Evidences in Citation-Based Search Engines: Is Microsoft Academic Search dead?

Enrique Orduña-Malea[1], Juan Manuel Ayllón[2], Alberto Martín-Martín[2], Emilio Delgado López-Cózar[2]

[1] EC3: Evaluación de la Ciencia y de la Comunicación Científica, Universidad Politécnica de Valencia (Spain)
[2] EC3: Evaluación de la Ciencia y de la Comunicación Científica, Universidad de Granada (Spain)

**ABSTRACT**

The goal of this working paper is to summarize the main empirical evidences provided by the scientific community as regards the comparison between the two main citation-based academic search engines: Google Scholar (GS) and Microsoft Academic Search (MAS), paying special attention to the following issues: coverage; correlations between journal rankings; and usage of these academic search engines. Additionally, self-elaborated data is offered, which are intended to provide current evidence about the popularity of these tools on the Web, by measuring the number of rich files (PDF, PPT and DOC) in which these tools are mentioned, the amount of external links that both products receive, and the search queries' frequency from Google Trends. The poor results obtained by MAS led us to an unexpected and unnoticed discovery: Microsoft Academic Search is outdated since 2013. Therefore, the second part of the working paper aims at advancing some data demonstrating this lack of update. For this purpose we gathered the number of total records indexed by MAS since 2000. The data shows an abrupt drop in the number of documents indexed from 2,346,228 in 2010 to 8,147 in 2013. This decrease is offered according to 15 thematic areas as well. In view of these problems it seems logical not only that MAS was poorly used to search for articles by academics and students (who mostly use Google or Google Scholar), but virtually ignored by bibliometricians.

## 1. OBJECTIVES

At the beginning of the second decade of the XXI century, the two major academic search engines with information about scientific citation were Google Scholar (GS) and Microsoft Academic Search (MAS), developed by two companies (Google and Microsoft), rivaling not only in the design of these tools but in a wide range of products and web services, being especially important for our research area the competition between their search engines (Google and Bing).

The goal of this working paper is to summarize the main empirical evidences provided by the scientific community as regards the comparison between these two products, compiled from the most relevant and recent works so far. Thus, the following data are available:

a) Coverage: number of documents indexed, number of citations retrieved, size of diverse bibliometric indicators (such as h-index, g-index, etc.).
b) Correlations between journal rankings.
c) Usage of GS and MAS by one small sample of scientists specialized in bibliometrics, and another sample of students and academics.

Additionally, self-elaborated data are offered, which are intended to provide current evidence about the popularity of these tools on the Web, by measuring:

a) The amount of rich content documents (PDF, PPT and DOC) in which these tools are mentioned: page count indicator.
b) The amount of external links that both products receive: web visibility indicator, and
c) The global trends of "Microsoft Academic Search", "Google Scholar", and "Google Scholar Citations" search queries.

The comparative analysis of the performance in the search, treatment, download, management and visualization of the information provided by the interfaces of these two search engines are excluded. These issues have been addressed in great detail by Jacsó (2011; 2012).

## 2. DATA SOURCE

Data on coverage, correlation and usage of Google Scholar and Microsoft Academic Search have been collected from the following documents:

- "Coverage and adoption of altmetrics sources in the bibliometric community" (**Haustein** et al, 2014).
- "How readers discover content in scholarly journals. Comparing the changing user behaviour between 2005 and 2012 and its impact on publisher web site design and function" (**Gardner** and **Inger**, 2013).
- "Microsoft Academic Search and Google scholar citations: comparative analysis of author profiles" (**Ortega** and **Aguillo**, 2014).
- "The number of scholarly documents on the public web" (**Khabsa** and **Giles**, 2014).

- "Ranking top economics and finance journals using Microsoft Academic Search versus Google Scholar: how does the new publish or perish option compare?" (**Haley**, 2014).

Otherwise, the sources used (as of April 2014) for the web popularity of Google Scholar and Microsoft Academic Search the Web are set below:

- *Search engines*: Google and Bing.
  http://google.com
  http://bing.com
- *Citation-based academic search engines*: Google Scholar (GS), Google Scholar Citations (GSC), and Microsoft Academic Search (MAS).
  http://scholar.google.com
  http://scholar.google.com/intl/en/scholar/citations.html
  http://academic.research.microsoft.com
- *Query trends*: Google Trends.
  http://www.google.com/trends
- *Link sources*: MajesticSEO, OpenSiteExplorer and Ahrefs.
  http://www.majesticseo.com
  http://www.opensiteexplorer.org
  https://ahrefs.com

## 3. RESULTS

### 3.1. Google Scholar versus Microsoft Academic Search

First, data on the coverage of GS and MAS and the correlation between rankings produced by both tools are outlined; after this, data on the usage of these tools on the Web, both from the perspective of the user seeking information about the tools, and from the view of the content creator who mention these products are presented.

### a) Coverage & rankings

The first work to be reviewed (**Haley**, 2014), compares the bibliometric performance of 50 top economics and finance journals both in GS and MAS using the Publish or Perish (PoP) application (**Harzing**, 2007). Two different time frames were tried: over "entire life" span in the target databases and the "1993-2012" time frame. Data were collected in June 2013.

The results obtained by the authors are clear and definite: GS doubled -and in some cases tripled- bibliometric values of all the indicators used to determine the impact of the 50 top economics and finance journals studied (Table 1).

Table 1. Bibliometric indicators of top 50 economics and finance journals[*]

| INDICATOR | ALL | | 1993-2012 | |
|---|---|---|---|---|
| | GS | MAS | GS | MAS |
| h-index | 196 | 90 | 154 | 79 |
| g-index | 352 | 148 | 267 | 126 |
| AWCR | 11,069 | 2,978 | 9,834 | 2,741 |
| e-index | 251 | 98 | 186 | 81 |

* Data source: re-elaborated from Haley (2014)

Another aspect analysed in this work is the degree of correlation between the rankings of journals elaborated from GS and MAS. Spearman rank tests were used to compare the rankings. The results obtained (Table 2) show a good correlation in all indices used, being higher when the timeframe is restricted to more recent years.

Table 2. Spearman correlation rank tests for top 50 economics and finance journals[*]

| INDICATOR | ALL | 1993-2012 |
|---|---|---|
| h-index | 0.772 | 0.890 |
| g-index | 0.763 | 0.843 |
| AWCR | 0.846 | 0.892 |
| e-index | 0.777 | 0.838 |

*Data source:* re-elaborated from Haley (2014)

Nonetheless, this work suffers from two methodological weaknesses that may influence the results:

- Inaccurate search queries of journal titles due to the nonuse of neither all possible variants of a journal name nor the "exclusion operator" to remove non-relevant documents.
- The existing limitation in GS to show only top 1000 results prevents from having the certainty that the results are valid since most of the searches performed for the 50 journals analysed far exceed the threshold set by GS.

The difference detected in the speed of data processing from searches conducted on MAS and GS are due to the difficulties imposed on GS massive searches involving PoP, which has prevented the realization of bibliometric studies that require many searches and lots of data.

The second paper to review (**Ortega & Aguillo**, 2014) offers a comparative analysis of the personal profiling capabilities of MAS and GSC. It should be specified to properly interpret the results that this paper does not offer a comparison between GS and MAS but for the author profiles provided by Google Scholar Citations (GSC) and those offered by MAS (specifically, 771 personal profiles appearing in both GSC and MAS databases).

The main results of that research clearly show that:

1. The number of profiles in MAS is almost 200 times the current number of profiles in GSC. MAS contained 19 million author profiles in August 2012. In the case of GSC, that information is unknown, but the authors estimate that in June 2012 the figure reached 106,246. The reason for this remarkable difference is the way in which both products are made: profiles in MAS are automatically created whereas the GSC profiles are created only when the end user (hopefully an author) freely decides to do it (and publicly to display it, because there are many profiles created but not published).

2. Microsoft Academic Search has two serious technical problems. On one hand a higher number of duplicated profiles, especially in languages with many possible name variants and different translations (such as Spanish, Portuguese, Chinese, and Russian) and on the other hand a lower updating rate (41% of the MAS profiles presented an outdated affiliation). A specific crawl of Stanford University's profiles indicated that 22% of the profiles had been inactive since the year 2000.

3. The main weakness of GSC is linked to its own product design: the profiles are to be created by the researchers (the authors are free to include information [professional affiliation, keywords], and the documents they wish). This also introduces a bias in behalf of authors strongly linked with the new information technologies, and it may cause the intentional manipulation of indicators (**Delgado**, **Robinson-García** & **Torres-Salinas**, 2014).

4. The MAS profiles are in general disciplinarily better balanced whereas GSC shows a strong bias toward the information and computing sciences.

In any case, we believe that the main contribution of the study is derived from the analysis of the 771 personal profiles appearing in both the MAS and the GSC. This comparison allows us to know what the differences in the number of documents and citations provided by each product are. The results speak for themselves: GSC gathers 158.3% more documents per profile than MAS, 327.4% more citations, and 155.8% higher h-index values. These differences occur in virtually every scientific discipline (Figures 1 and 2) except for Chemistry and Medicine. However, it is striking that in these two disciplines MAS gathers more documents than GSC but recovers far fewer citations. This contradiction is surprising and it might have been caused because the sample taken in these areas was not big enough.

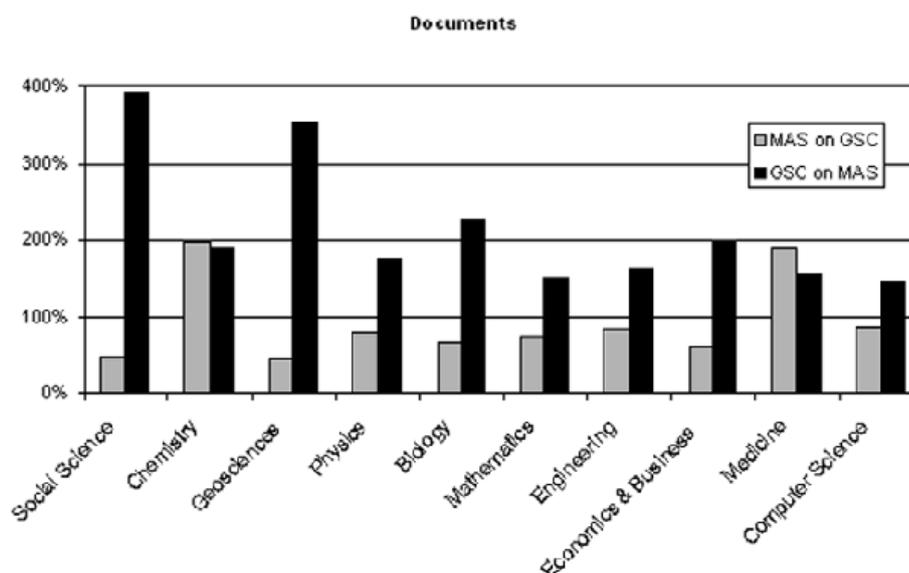

**Figure 1. Proportion of documents gathered by GSC and MAS according to disciplines**
Data source: Ortega & Aguillo (2014)

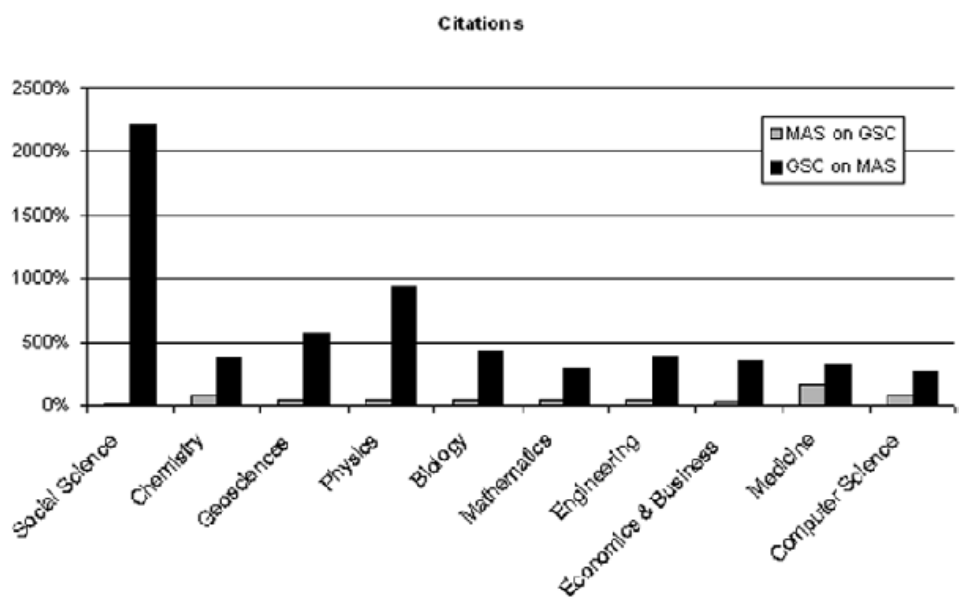

**Figure 2. Proportion of citations gathered by GSC and MAS according to disciplines**
Data source: Ortega & Aguillo (2014)

As an illustrative example, the differences that GSC and MAS exhibit for four of the bibliometric researchers cited in this report are showed in Table 3.

**Table 3. Metric values of some bibliometricians in MAS and GS**

| AUTHORS | DOCUMENTS | | CITATIONS | |
|---|---|---|---|---|
| | MAS | GS | MAS | GS |
| **Isidro F Aguillo** | 56 | 261 | 169 | 1,531 |
| **Jose Luis Ortega Priego** | 45 | 60 | 125 | 841 |
| **Peter Jacsó** | 84 | 494 | 479 | 2,498 |
| **Judit Bar-Ilan** | 99 | 145 | 1,145 | 3,763 |

Data source: self-elaborated

Finally, a new recent study intends to calculate the number of English scholarly documents on the public web (**Khabsa** and **Giles**, 2014), from a sample of 150 documents in 15 different topic areas (which corresponds with MAS thematic classification), and applying the Lincoln-Petersen method (the same employed by **Ortega** and **Aguillo** while estimating the number of personal profiles) to infer the English scholarly public web (which is considered as the summation of MAS and GS databases, avoiding the overlap between both search engines).

For the 150 documents, the authors found 86,870 citations from GS, and only 41,778 from MAS. Moreover, the Google Scholar database is estimated to have 99.3 million documents, compared with the near 50 million documents of MAS (in this case, this figure is taken from the information offered by Microsoft Azure Marketplace), although this figure may be slightly underestimated due to the method employed. This issue will be discussed in a separate working paper.

Figure 3 shows the relative number of documents by search engine:

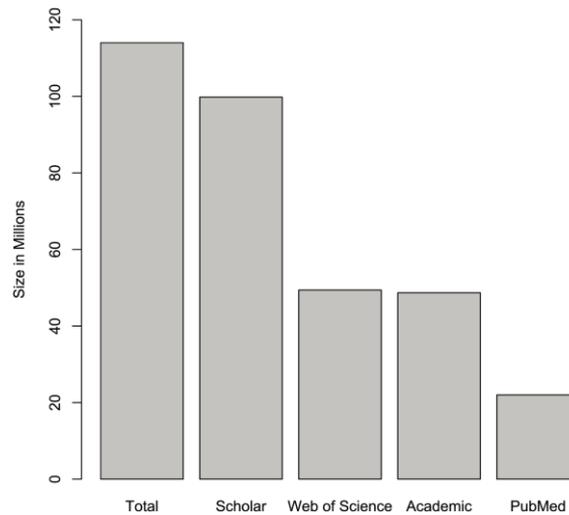

**Figure 3. Size of different academic search engines and databases**
Data source: Khabsa & Giles (2014)

The authors also provide information of the size according to the 15 disciplines considered (Figure 4), where we can observe the superiority of Google Scholar especially on Multidisciplinary, Social Sciences, Arts & Humanities, and Physics. These data (avoiding the differences of discipline classification, method used, and date of data collection) are similar of those obtained by Ortega & Aguillo (shown previously on Figure 1), except for:
- Medicine: Ortega & Aguillo put MAS over GS, as mentioned before.
- Computer Sciences: Khabsa & Giles put MAS over GS.

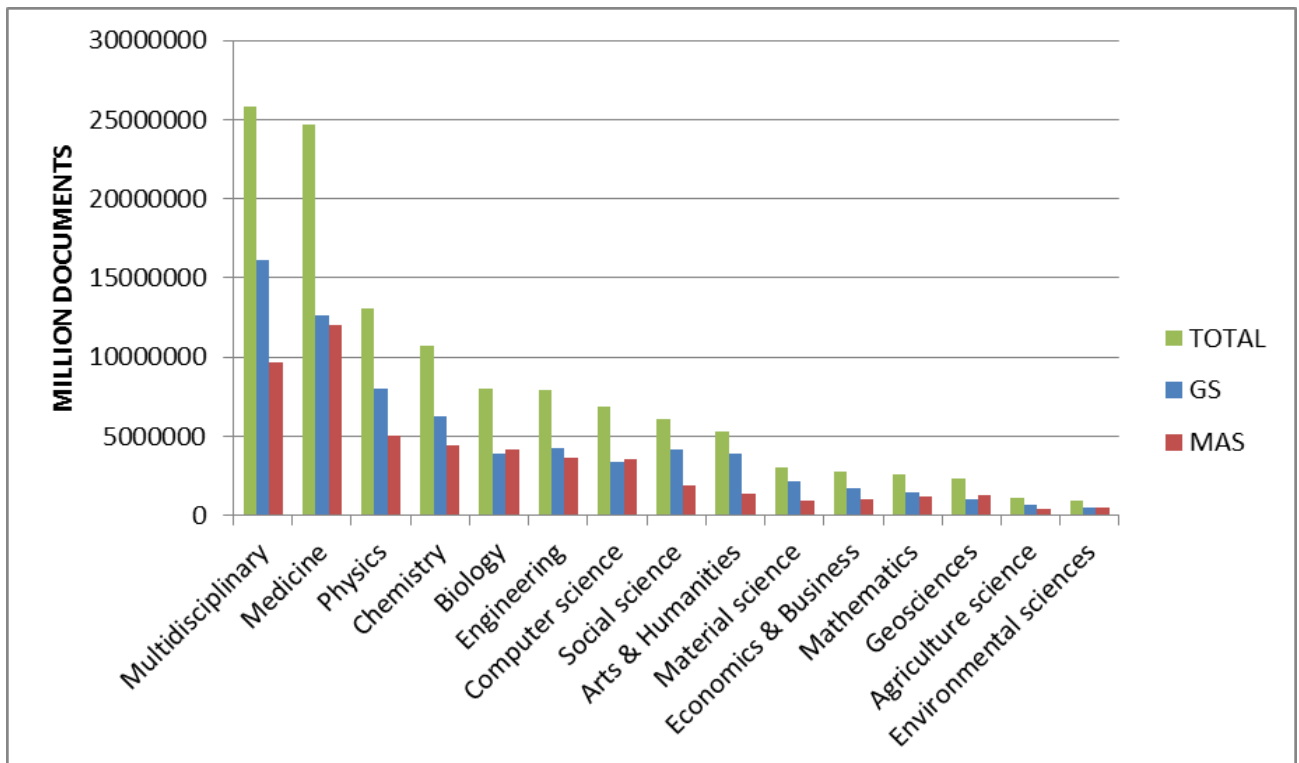

**Figure 4. Size of different academic search engines and databases (Lincolm-Petersen)**
Data source: re-elaborated from Khabsa & Giles (2014)

**b) Degree of use**

In order to know to what extent GS and MAS are used, we recovered two empirical studies -although not primarily aimed to perform a comparative analysis of these products- which provide data on them, based on surveys.

The first, performed by **Gardner** and **Inger** (2013), seeks to learn how readers discover, access and navigate in the content of scholarly journals. This is a large scale survey of journal readers (n = 19,064) conducted during May, June and July of 2012. All regions of the world and all professional sectors, especially the academic researchers (50% respondents) and students (20% of respondents) are well represented. Therefore it aptly reflects academia worldwide.

This study concludes that "when searching and following a citation, academic search engines are the second most popular resource across the board. Instead, they are less important for people who want to discover latest articles". For our analysis, the most interesting question that this study answered was: "What are the users' preferred search engines to seek Journal articles?" (Figure 5). The data are overwhelming: Google and Google Scholar are always the first choice (the latter especially by students). Microsoft Academic Search is not practically used.

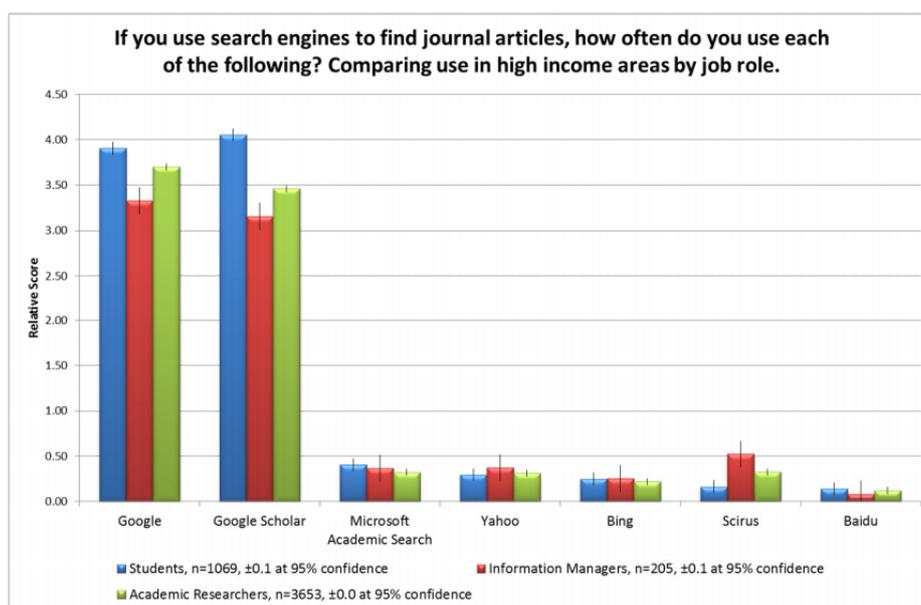

**Figure 5. Preferred search engines for users to seek journal articles**
Data source: Gardner & Inger (2013)

The second work (**Haustein** et al 2014), was addressed to a highly specialized but also very qualified sample: the bibliometricians. This research intends to determine the use and coverage of social media environments, examining both their own use of online platforms and the use of their papers on social reference managers. The survey was distributed among the 166 participants (71 returned the questionnaire) of the 17th International Conference on Science and Technology Indicators (STI) in Montréal from September 5th to 8th, 2012.

As the authors state: "asked for personal publication profiles on Academia.edu, Google Scholar Citations, Mendeley, Microsoft Academic Search, ResearcherID (WoS), or ResearchGate, 32 participants listed their publications at least at one of these platforms. The most popular tool was Google Scholar Citations (22 respondents with profile; 68.8% of those with publication profiles)". MAS is the second least used platform, at a considerable distance from Google Scholar Citation, as may be observed in Table 4.

Table 4. Bibliometricians with author profiles in various social platforms*

| ACADEMIC SOCIAL PLATFORM | NUMBER OF AUTHORS WITH PROFILE |
|---|---|
| Google Scholar Citations | 22 |
| Researcher ID | 14 |
| ResearchGate | 9 |
| Mendeley | 8 |
| Microsoft Academic Search | 7 |
| Academia.edu | 5 |

* Data source: Haustein et al. (2014)

When bibliometricians were asked what they were doing with their publication profiles (Figure 6), the authors found that GSC was the most used in all typical activities related with the maintenance of an author profile: check citations, add missing publications, and merge same publications; and especially mostly used to check citations. However, MAS was the least used in all operations. Noteworthy, the people especially used this service to delete "wrong" publications from their profiles. This would confirm the technical problems of MAS, subsequently detected by Aguillo & Ortega (2014).

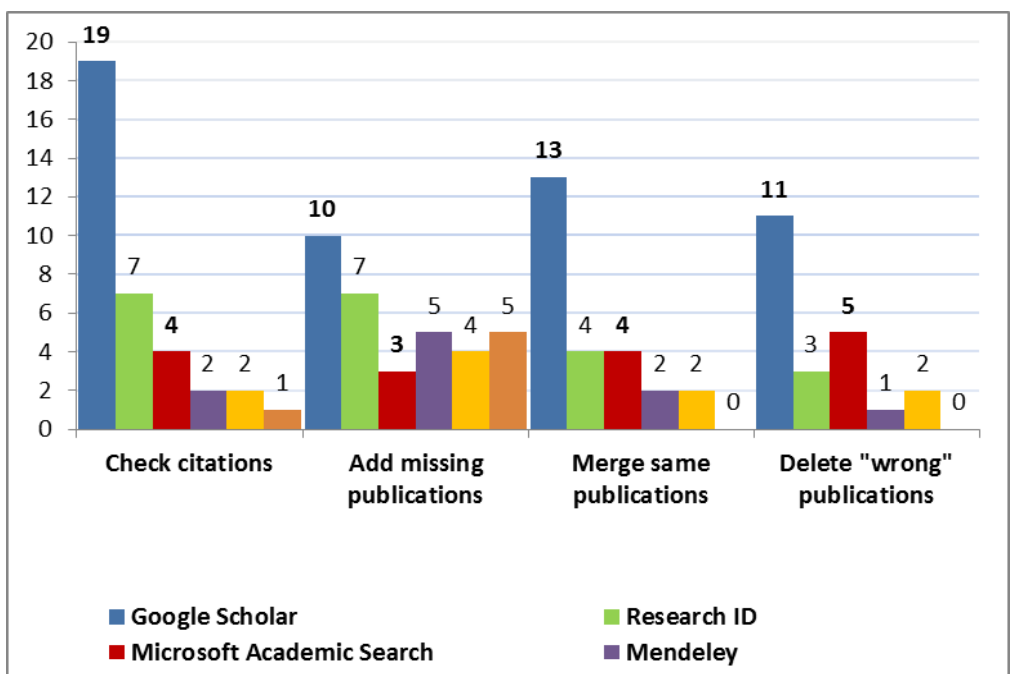

Figure 6. Use of Author profiles in various platforms by a bibliometricians sample
Data source: Haustein et al. (2014)

The data collection was repeated in November 2013, showing an exponential growth of GSC profiles. The percentage of researchers with GSC profiles

**9**

increased from 23% in February 2012 to 53% in November 2013, thereby indicating the popularity of this platform.

**c) Visibility of GS and MAS websites through webmetric measures**

To complete this section we have performed three brief analyses that aim to investigate indirectly the use and popularity of GS as MAS on the web space.

*Web presence: number of title mentions*

First, we searched for documents on the web that refer to GS and MAS (Table 5). We limited the search to those rich content document file types (PDF, PPT, and DOC), the most widely used in scientific and academic environments, and excluding the references made within their corresponding web domains (ie., google.com and microsoft.com).

The queries were submitted both to Google and Bing search engines, using the following search commands[1]:
<"Google Scholar" -site:google.com filetype:xxx>
<"Microsoft Academic Search" -site:microsoft.com filetype:xxx>

Table 5. Number of documents (PDF / PPT / DOC files) related to "Google Scholar" and "Microsoft Academic Search" gathered from Google (April, 2014)

| FILE TYPE | GOOGLE | | BING | |
|---|---|---|---|---|
| | GS | MAS | GS | MAS |
| PDF | 1,800,000 | 46,300 | 55,500 | 1,520 |
| DOC | 6,090 | 32 | 5,390 | 37 |
| PPT | 3,740 | 30 | 3,839 | 31 |

Data source: self-elaborated

Despite the well-known differences between Google and Bing as regards the coverage of the Web (which are outside the scope of this working paper), the differences between mention figures in both search engines are similar (especially for DOC and PPT files). The results, again, are extremely enlightening. Considering all three file types, the web presence of "Google Scholar" is overwhelmingly superior to "Microsoft Academic Research", both in Google and Bing search engines.

*Web visibility: number of URL mentions, backlinks and sites linking*

The web visibility of the corresponding web domains of GS and MAS have been checked by means of three different sources (Open Site Explorer, MajesticSEO and Ahrefs) in order to check the number of external links and especially -due to its importance- the root domains (number of sites linking in) referring to the academic search engines under analysis (Table 6)[2].

---

[1] The expression "xxx" were substituted for "pdf", "doc" and "ppt" in each search query correspondingly.
[2] MAS was named "Academic Live Research" from 2006 to 2008, and accessed by the following URL: <http://academic.live.com>. In any case, at present this website and name is not representative.

The results clearly state the higher visibility of GS over MAS in all three sources, regardless of considering either external links or referring domains, and despite the different link coverage of each platform.

Table 6. Web visibility of GS and MAS in various link sources

| SEARCH ENGINE | URL | | OSE | MAJESTIC SEO | AHREFS |
|---|---|---|---|---|---|
| GS | scholar.google.com | External Links | 12,300,958 | 78,604,588 | 183,000,000 |
| | | Referring Domains | 34,137 | 61,759 | 62,000 |
| MAS | academic.research.microsoft.com | External Links | 2,310,927 | 13,507,218 | 18,000,000 |
| | | Referring Domains | 2,368 | 4,585 | 4,400 |

Data source: self-elaborated

### Popularity: number of search queries

Finally, we used the Google Trends service in order to identify which is the most popular academic search engine in search queries made by users. We compared GSC and MAS (Figure 7a), and MAS and GS (Figure 7b). The differences in favor of Google products are awesome. For GSC, as can be seen in Figure 5a, the user queries have not stopped growing since its birth, rapidly beating MAS, which shows a progressive decline.

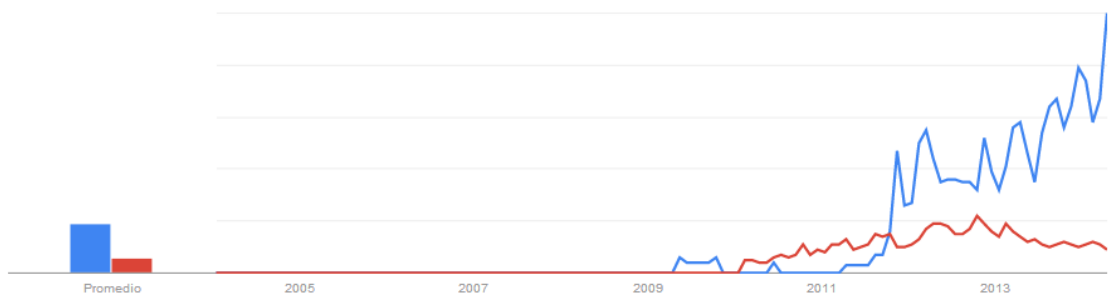

**Figure 7a. "Google Scholar Citations" & "Microsoft Academic Search" search queries**[*]

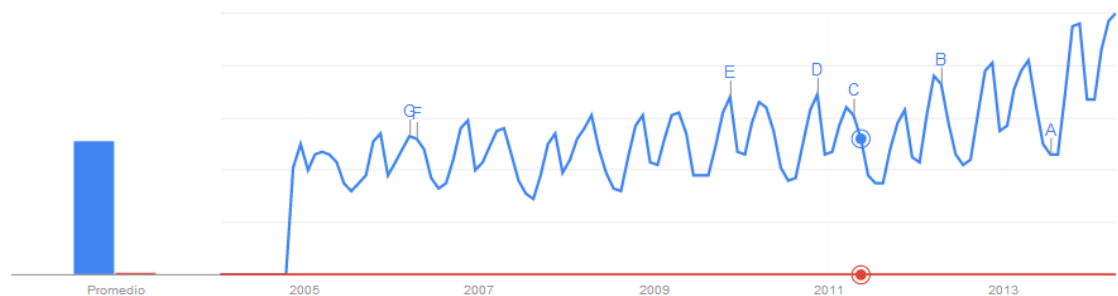

**Figure 7b. "Microsoft Academic Search" & "Google Scholar" search queries**[*] Data source: Google Trends <google.com/trends>

**11**

### 3.2. Microsoft Academic Search update

Although the initial purpose of this report was to present available empirical evidence to compare the only two existing citation-based academic search engines today (Google Scholar and Microsoft Academic Search), the excessively poor results obtained for MAS in the previous sections led us to an unexpected discovery that altered our initial goal: the lack updates of Microsoft Academic Search data from 2013; process that had begun in 2011, when coverage plummeted.

Therefore, in this second part of the working paper, our objective is to advance some data demonstrating the lack of updates of Microsoft Academic Search in the last few years, waiting for a larger work in progress, where this fact will be noted and measured, and where its scope will be demonstrated.

In order to obtain indicative data that will allow us to reliably assess the degree of the search engine update, we gathered the number of total records indexed by MAS since 2000 (ie., including retrospective data). The results are displayed in Table 7.

Table 7. Evolution of the number of publications indexed in MAS (2000-2014)

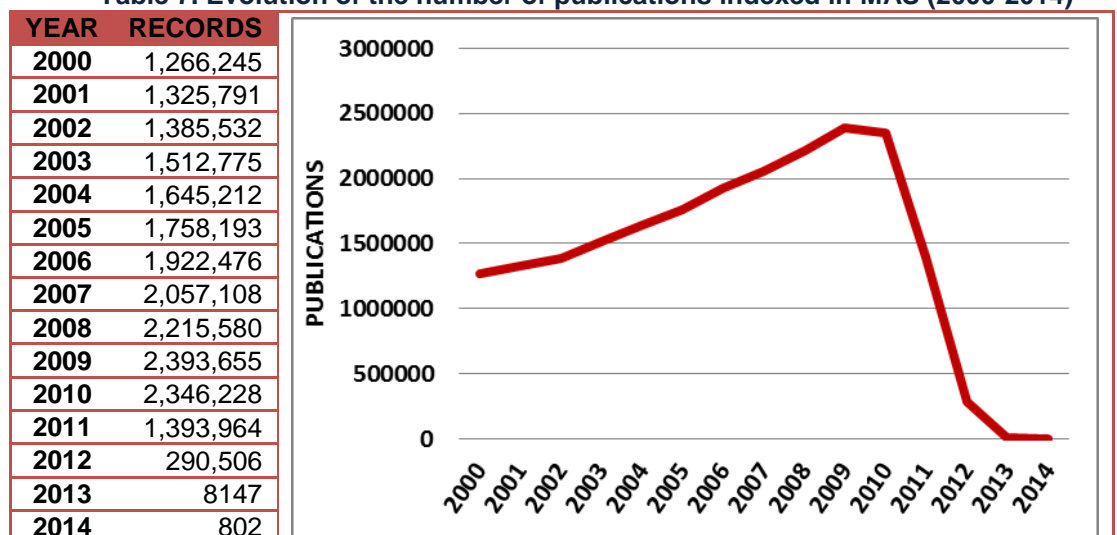

| YEAR | RECORDS |
|---|---|
| 2000 | 1,266,245 |
| 2001 | 1,325,791 |
| 2002 | 1,385,532 |
| 2003 | 1,512,775 |
| 2004 | 1,645,212 |
| 2005 | 1,758,193 |
| 2006 | 1,922,476 |
| 2007 | 2,057,108 |
| 2008 | 2,215,580 |
| 2009 | 2,393,655 |
| 2010 | 2,346,228 |
| 2011 | 1,393,964 |
| 2012 | 290,506 |
| 2013 | 8147 |
| 2014 | 802 |

Data source: self-elaborated

The data presented in Table 7 shows a period of higher indexation (from 2007 to 2010), a significant drop in 2011 and alarming hereinafter.

Surprisingly, it seems that this fact has gone unnoticed by the community, probably due to the lack of real use, as data in the previous section proved. This issue highly contrasts with the available Microsoft Academic Search API via the Windows Azure Marketplace[3], with the announced information of new features in January 2013 (more than 10 million new publications from JSTOR, Nature, PLoS, SSRN, and 23 additional publishers added)[4], or the manual inclusion of new journals in 2014, as showed in the official online forum.[5]

---

[3] http://datamarket.azure.com/dataset/mrc/microsoftacademic
[4] http://academic.research.microsoft.com/About/Help.htm
[5] http://social.microsoft.com/Forums/en-US/home?forum=mas

Moreover, the data about the size of MAS is contradictory:

a) Data shown on the web:
http://academic.research.microsoft.com/About/Help.htm

> As for September 2011: "The number of publications increases to 35.3 million".
>
> As for January 2013: "More than 10 million new publications from JSTOR, Nature, Public Library of Science (PLoS), SSRN, and others (23 publishers added)".

So we can estimate at least 45.3 million publications. In fact, Khabsa and Giles assume the size of MAS in 48,774,763 documents in their estimation of the English scholarly public web.

b) Data shown in Azure: 39.85 million documents (table name: paper).
http://datamarket.azure.com/dataset/mrc/microsoftacademic

Moreover, the 10 million new documents added in 2013 do not match with the figures shown in this working paper. Part of this documents may be retrospective data (or not updated yet since January 2013).

Nonetheless, the data offered in Table 7 is incontestable. To make sure that the upgrade issues are general and that there are no differences according to scientific disciplines, the raw data has been divided into the 15 thematic areas that MAS uses to visualize the scientific information (Figure 8): Agriculture Science, Arts & Humanities, Biology, Chemistry, Computer Science, Economics & Business, Engineering, Environmental Sciences, Geosciences, Material Science, Mathematics, Medicine, Multidisciplinary, Physics, Social Science.

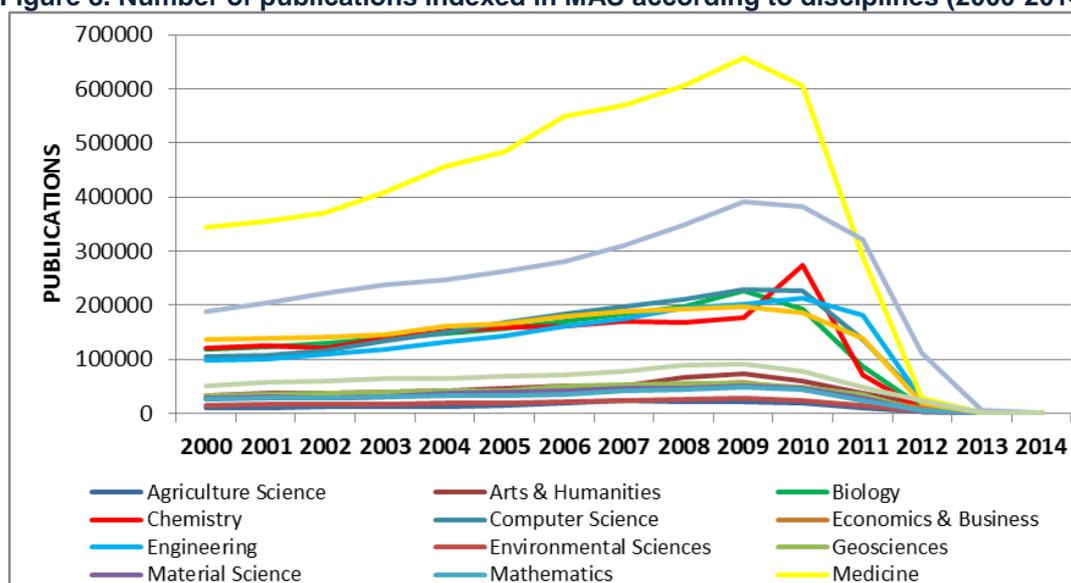

**Figure 8. Number of publications indexed in MAS according to disciplines (2000-2014)**

Data source: self-elaborated

Even in "Medicine" and "Multidisciplinary" (which covers the journals Nature and Science, among others), the areas with a higher output indexed, the drop is

**13**

abrupt since 2011, with an inexplicable slowdown in "Multidisciplinary", exemplified by Science and Nature performance charts, powered by MAS (Figure 9).

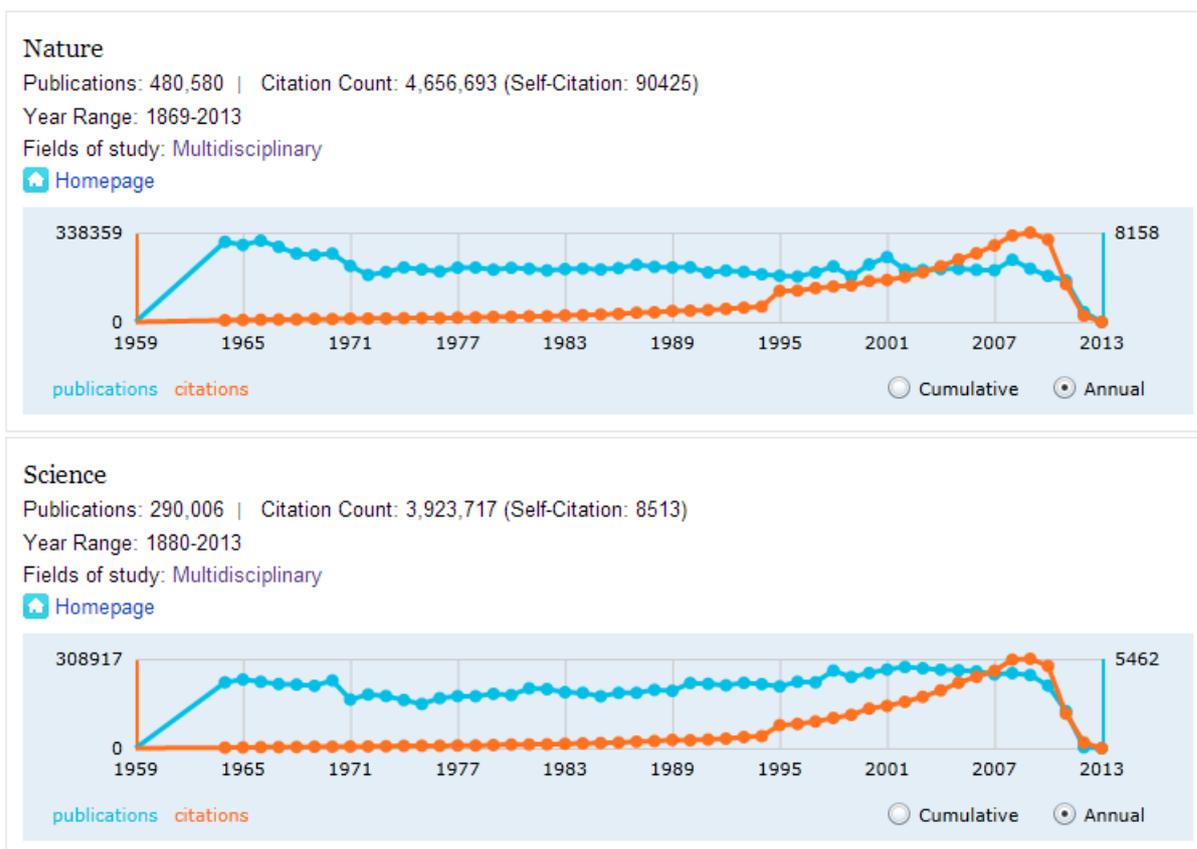

**Figure 9. Nature and Science performance charts in Microsoft Academic Search**
Data source: Microsoft Academic Search

## 4. CONCLUSIONS

The first and unexpected result of this report is that MAS has no longer been updated since 2013, although this phenomenon began to be glimpsed in 2011, when its coverage plummeted. This issue has gone unnoticed, as far as we know, in the bibliometric and webometric arena.

Analysing the results of the empirical evidence comparing MAS with GS, it's no wonder the collapse of MAS, since GSC contains more academic materials that produce more citations than MAS. The GS services doubles, and even triples, MAS values, both when a small sample of journals, or a set of authors' bibliometric profiles are compared over the two products. Moreover, the empirical evidence studies also noted that MAS had low updating rates and contained a lot of duplicate information (multiple profiles of the same author).

In view of these problems it seems logical not only that MAS was poorly used to search for articles by academics and students (who mostly use Google or Google Scholar), but virtually ignored by bibliometricians, who choose Google Scholar Citations for keeping their public profile updated and for periodically checking citations received.

The ultimate proof of the broad acceptance that Google Scholar and Google Scholar Citations have is the higher number of rich content documents that contain references to those products, the higher number of links received as web targets or the higher number of search queries performed looking for this topic. All these data indicate that Google products are not only used but also taught, becoming objects of research and reflection as well. However, MAS was virtually ignored by users. Even its disappearance has been ignored, although the activity of official forums and inclusion of new journals in 2014 should be further analysed in order to better explain what is really happening with the product.

Finally, we can only be saddened by the loss of a product such as MAS, which poked healthy competition and had deployed smart visualizing tools. We do hope that Microsoft Academic Research can rise back for the sake of scientific evaluation.